\documentclass[aps,prd,reprint,superscriptaddress,showpacs]{revtex4-1}
\pdfoutput=1
\usepackage{hyperref}
\usepackage{graphicx}
\usepackage{enumerate}
\hypersetup{
    pdfnewwindow=true,      
    colorlinks=true,       
    linkcolor=blue,          
    citecolor=blue,        
    filecolor=blue,      
    urlcolor=blue           
}

\usepackage{subfigure}
\usepackage{amsmath}
\usepackage{amssymb}
\usepackage{amsfonts}
\usepackage{color}

\def\aj{Astron.\ J.}
\def\apj{Astrophys.\ J.}
\def\apjl{Astrophys.\ J. Lett.}

\def\mnras{Mon.\ Not.\ Roy.\ Astron.\ Soc.}

\def\prl{Phys.\ Rev.\ Lett.}
\def\prd{Phys.\ Rev.\ D}

\def\be{\begin{equation}}
\def\ee{\end{equation}}

\begin{document}

\title{Probing the Structure of Jet Driven Core-Collapse Supernova and \\Long Gamma Ray
  Burst Progenitors with High Energy Neutrinos}

\author{Imre Bartos}
\email{ibartos@phys.columbia.edu}
\affiliation{Department of Physics, Columbia University, New York, NY 10027, USA}
\author{Basudeb Dasgupta}
\email[]{dasgupta.10@osu.edu}
\affiliation{\mbox{Center for Cosmology and AstroParticle Physics, Ohio State University, Columbus, OH 43210, USA}}
\author{Szabolcs M\'arka}
\affiliation{Department of Physics, Columbia University, New York, NY 10027, USA}

\def\be{\begin{equation}}
\def\ee{\end{equation}}

\newcommand{\todo}{\textcolor{red}}

\begin{abstract}
Times of arrival of high energy neutrinos encode information about their sources. We demonstrate that the energy-dependence of the onset time of neutrino emission in advancing relativistic jets can be used to extract important information about the supernova/gamma-ray burst progenitor structure. We examine this energy and time dependence for different supernova and gamma-ray burst progenitors, including red and blue supergiants, helium cores, Wolf-Rayet stars, and chemically homogeneous stars, with a variety of masses and metallicities. For choked jets, we calculate the cutoff of observable neutrino energies depending on the radius at which the jet is stalled. Further, we exhibit how such energy and time dependence may be used to identify and differentiate between progenitors, with as few as one or two observed events, under favorable conditions.
\end{abstract}


\maketitle

\section{Introduction}

There is growing observational evidence and theoretical foundation for the connection between core-collapse supernovae (CCSNe) and long gamma-ray bursts (GRBs)~\cite{1998Natur.395..670G, bloom:02, 2003ApJ...591L..17S}. Location of GRBs in blue-luminosity regions of their host galaxies, where massive stars form and die, and CCSN signatures in the afterglows of nearby GRBs have provided strong evidence for the CCSN-GRB connection (see summaries by \citep{wb:06,2011arXiv1104.2274H,modjaz:11}). While many details of the CCSN-GRB relationship are still uncertain, current theoretical models suggest that a canonical GRB
has a relativistic jet from a central~engine~\cite{Cavallo:1978zz, Goodman:1986az, Paczynski:1986px, Piran:1999kx, Zhang:2003uk, Piran:2004ba, 2006RPPh...69.2259M}.

Gamma rays are emitted by high energy electrons
in the relativistic jet~\citep{Fireball}. If the relativistic jet is able to escape from the star, the gamma rays can be observed and the GRB is coined ``successful.'' If the jet stalls inside the star, however, the gamma-ray signal is unobservable and the GRB is coined ``choked''~\citep{2001PhRvL..87q1102M}. While only $\sim 10^{-3}$ of CCSNe has extremely relativistic jets with $\Gamma\gtrsim 100$, which lead to successful GRBs, a much larger subset of non-GRB CCSNe appears to be accompanied by collimated mildly relativistic jets with $\Gamma\sim 10$~\cite{FailedGRBandorphan2002MNRAS.332..735H, soderberg:04, 2006Natur.442.1011P, modjaz:11, 2012MNRAS.420.1135S, 2010Natur.463..513S}. These CCSNe with mildly relativistic jets may make up a few percent of all CCSNe \cite{berger:03, soderberg:04, vanPutten:2004dh, HeEnvelopeBreakout}. Similarly, the jet's Lorentz factor for low-luminosity GRBs may also be much below those of high luminosity GRBs~\mbox{\citep{2006ApJ...651L...5M,2007APh....27..386G,2007PhRvD..76h3009W, Baret20111}}.

GRB progenitors may also produce high energy neutrinos (HENs) \cite{2010RScI...81h1101H}. The relativistic jet responsible for the gamma rays has been argued to be responsible for shock-acceleration of protons to ultrarelativistic energies, leading to nonthermal HENs produced in photomeson interactions of the accelerated protons.
Recent upper limits from the IceCube detector \citep{2004APh....20..507A} disfavor GRB fireball models with strong HEN emission associated with cosmic ray acceleration. However, milder HEN fluxes or alternative acceleration scenarios are not ruled out \cite{2012Natur.484..351A}. Moreover, the constraints weaken substantially when uncertainties in GRB astrophysics and inaccuracies in older calculations are taken into account, and the standard fireball picture remains viable \cite{PhysRevLett.108.231101,2012ApJ...752...29H}.
Proton acceleration may occur in external shocks \cite{Fireball, Murase:2007yt}, both forward and reverse, as well as in internal shocks associated with the jet~\cite{Narayan:1992iy, Rees:1994nw, waxmanbachall, Rachen:1998fd, Waxman:1999ai, Li:2002dw, Dermer:2003zv}.
HENs, unlike gamma rays, can thus be emitted while the jet is still inside the star. Therefore, while gamma rays are expected to be emitted only if the GRB is successful, HENs are expected for both successful and choked GRBs. Thus, not only does the expected rate increase to a larger fraction of the CCSN rate of $(2-3)\,{\rm yr}^{-1}$ in the nearest $10\,{\rm Mpc}$ \cite{Ando:2005ka, Horiuchi:2011zz, Botticella:2011nd}, these mildly relativistic jets also present a more baryon-rich environment conducive to neutrino emission. Emission of HENs from internal shocks of mildly relativistic jets has therefore been considered to be an important contribution to the overall observable neutrino flux \cite{HEN2005PhysRevLett.93.181101, HENAndoPhysRevLett.95.061103, chokedfromreverseshockPhysRevD.77.063007}.

As only neutrinos and gravitational waves can escape from the inner regions of a star, it means that they are a unique tool to study the internal structure of GRB progenitors. Whether and when HENs escape the star depend on the star's optical depth to neutrinos, which in turn depends on where neutrinos are emitted inside the star, as well as the neutrino energy -- HENs can only escape after the relativistic jet reaches a radius where densities are low enough. This finite neutrino optical depth can delay and modify the spectrum of HEN emission from GRBs. Detecting HENs with time and energy information, at present and upcoming neutrino telescopes, therefore presents an unprecedented opportunity to probe GRB and CCSN progenitors, which may shed light on the GRB mechanism and the CCSN-GRB connection.

Probing the interior structure of GRB/SN progenitors via HENs has first been suggested by Razzaque, M\'esz\'aros, and Waxman \cite{PhysRevD.68.083001}. They showed, for two specific stages of jet propagation, that the observable neutrino spectrum is affected by the stellar envelope above the jet head, which can in turn be used to examine this envelope via the detected neutrinos. Horiuchi and Ando \citep{chokedfromreverseshockPhysRevD.77.063007} also mention HEN
interactions for jets within the stellar envelope (see also Section \ref{section:interaction}).

This article examines the question: What do the times of arrival of detected high energy neutrinos tell us about the properties of their source? We investigate the role of the opacity of CCSN/GRB progenitors in the properties and distribution of observed HENs, and how these observed HEN properties can be used to probe the progenitors' structure. Studying the optical depth at which HENs can escape the progenitor, we find a progenitor- and energy-dependent temporal structure of the high energy neutrino emission and jet breakout. Observations of HEN signatures of CCSNe or GRBs at neutrino telescopes, even with one or two events, could provide crucial information for differentiating between progenitors and characterizing their properties. Such information would advance our understanding of CCSNe, GRBs, and their relationship to each other.

The paper is organized as follows: In Sec.~\ref{sec:progenitors}, we review CCSN and GRB models. In Sec.~\ref{sec:HENS}, we briefly discuss HEN production in GRBs, propagation in the stellar material and flavor oscillations thereafter, and detection at Earth. In Sec.~\ref{sec:optical}, we describe our calculations of the neutrino interaction length and optical depth inside the stellar envelope, and present our results on the energy-dependent radius from which neutrinos can escape. In Sec.~\ref{sec:temporal}, we discuss the temporal structure of energy-dependent neutrino emission from advancing and stalled jets for different stellar progenitors. This is followed in Sec.~\ref{sec:onsets} by our results for energy dependent onset and emission duration of HENs. In Sec.~\ref{sec:results}, we present our interpretations for the energy-dependent onset of HEN emission and discuss how it probes the progenitors, particularly with a few detected neutrinos. We summarize our results in Sec.~\ref{sec:summary}.

\section{CCSN and GRB Progenitors}
\label{sec:progenitors}

Current understanding of canonical long GRBs suggests that they are collapsars requiring a massive progenitor star that is (i) rapidly spinning \cite{woosley:93, Paczynski:1997yg, macfadyen:01, metzger:11} and (ii) has a small radius ($\sim$ solar radius) \cite{matzner:03, wb:06}. Successful GRBs also appear to prefer a lower metallicity \cite{Stanek:2006gc}, but choked GRBs may not require that. While this limited information does not always allow one to identify a specific progenitor, it does suggest that the progenitors are massive rotating stars \cite{wb:06}.

\emph{Rotating red and blue supergiants} (e.g., \cite{1989A&A...224L..17L,whw:02}) may be the progenitors of many GRBs. These stars are in the final stages of the pre-collapse evolution of massive stars, whose collapse can naturally lead to CCSNe and GRBs. Furthermore, some of these stars may lose their hydrogen envelope due to a binary companion, which can help the stars retain the fast rotation necessary for the creation of GRBs \cite{1998ApJ...494L..45P}.

\emph{Wolf-Rayet} (WR) stars are originally heavy, but lose their hydrogen envelope (and therefore a significant fraction of their mass) through stellar winds. A relativistic jet from a rotating WR star can therefore escape without having to penetrate a hydrogen envelope, making these stars a common type of progenitor \citep{whw:02,MNR:MNR18598}. Mass loss through stellar winds is expected to be significant for stars with higher metallicity \citep{2005A&A...442..587V}. One difficulty with such mass loss is that it carries away crucial angular momentum from the star. As the emergence of relativistic jets requires a very rapidly rotating core, losing angular momentum decreases the possibility of a GRB \citep{0004-637X-637-2-914}. As a result, compact progenitors (i.e. that have lost their hydrogen and/or helium envelopes) also regularly explode as type Ibc supernovae without indications of a central engine injecting jet power into the explosion (e.g., \cite{smith:11b}).

Alternatively to stellar winds, massive stars can lose their hydrogen envelope to a companion star, which can leave more angular momentum at the core. Such rotating objects, composed of the \emph{bare helium core} left behind, may also be GRB progenitors. Unusually rapid rotation on the main sequence can also result in mass loss \citep{0004-637X-637-2-914}.

Single stars with extremely rapid rotation may experience almost complete mixing on the main sequence \citep{0004-637X-637-2-914}, leading to a \emph{chemically homogeneous star}. Such stars bypass the red giant phase and resemble WR stars, but with little mass loss. This scenario is particularly interesting for GRB production as it combines low mass loss with rapid core rotation, the two prerequisites for GRB emission.

\begin{center}
  \begin{table}
\begin{center}
    \begin{tabular}{ l | c | c | c | c | c | c | c}
    \hline
    Model & $Z$         & $M$           & $M_{He}$         & $M_H$            & $R$              & $R_{He}$       & Ref. \\
          & $[Z_\odot]$ & $[$M$_\odot]$ & $[$M$_\odot]$    & $[$M$_\odot]$    & [10$^{13}$ cm]   & [$10^{11}$ cm] &      \\ \hline
    15L   & $10^{-4}$   & 14.9          & 4.5              & 7.4              & 0.4              & 1.4            & \citep{whw:02}    \\
    15Lc  & $10^{-4}$   & 5.2           & 1.7              & 0.4              & N/A              & 1.4            & \citep{whw:02}    \\
    15S   & 1           & 12.6          & 4.0              & 5.5              & 5.9              & 1.9            & \citep{whw:02}    \\
    16T   & $10^{-2}$   & 15.1          & 0.1              & $3\times10^{-4}$ & $8\times10^{-2}$ & N/A            & \citep{0004-637X-637-2-914} \\
    40S   & 1           & 8.7           & 0.1              & $10^{-3}$        & $8\times10^{-3}$ & 0.8            & \citep{whw:02}    \\
    75S   & 1           & 6.3           & 0.2              & $8\times10^{-4}$ & $7\times10^{-3}$ & 0.7            & \citep{whw:02}    \\ \hline
    12L   & $10^{-4}$   & 11.9          & 3.4              & 6.2              & 2.4              & 1.6            & \citep{whw:02}    \\
    20L   & $10^{-4}$   & 19.9          & 6.1              & 9.0              & 0.3              & 1.9            & \citep{whw:02}    \\
    25L   & $10^{-4}$   & 24.9          & 7.7              & 10.4             & 0.3              & 2.3            & \citep{whw:02}    \\
    35L   & $10^{-4}$   & 34.8          & 10.9             & 12.7             & 1.2              & 3.4            & \citep{whw:02}    \\
    12S   & 1           & 10.7          & 3.4              & 5.1              & 4.3              & 1.9            & \citep{Presupernova0004-637X-626-1-350} \\
    15Sb  & 1           & 11.9          & 3.8              & 4.8              & 6.1              & 2.3            & \citep{Presupernova0004-637X-626-1-350} \\
    20S   & 1           & 12.7          & 4.0              & 3.5              & 7.7              & 2.8            & \citep{Presupernova0004-637X-626-1-350} \\
    25S   & 1           & 12.2          & 3.2              & 1.7              & 8.2              & 3.2            & \citep{Presupernova0004-637X-626-1-350} \\
    35S   & 1           & 14.6          & 2.2              & 0.4              & 0.1              & 0.2            & \citep{Presupernova0004-637X-626-1-350} \\
    \hline
    \end{tabular}
\caption{Properties of pre-supernova stellar models used in the
  analysis. The columns are: model name, metallicity, pre-supernova
  (PS) stellar mass, helium mass, hydrogen mass, stellar radius,
  helium core radius, and reference to models. Model names contain the
  ZAMS stellar mass, and a letter representing metallicity (L - low, T
  - $1\%$ Solar, and S - Solar). There is additional differentiation
  between the various 15\,M$_\odot$ models we use.} \label{radiustable}
\end{center}
  \end{table}
\end{center}

Guided by these facts, we shall examine the following progenitor models in detail:
\begin{enumerate}
\item \emph{Red supergiant}        -- zero-age main sequence (ZAMS) mass of 15\,M$_\odot$, both with solar (15S) and low (15L) metallicities. We further study the 15L model with its hydrogen envelope removed (15Lc) due to, e.g., a companion star.
\item \emph{Wolf-Rayet star}             -- ZAMS mass of 75\,M$_\odot$ with solar metallicity (75S).
\item \emph{Bare helium core}            -- ZAMS mass of 40\,M$_\odot$ with solar metallicity (40S).
\item \emph{Chemically homogeneous star} -- ZAMS mass of 16\,M$_\odot$ with low metallicity (16T).
\end{enumerate}
We have indicated, in parentheses, the names of the models we consider in this study. To obtain the matter distribution and composition of these models, we use the numerical results of Woosley \emph{et al.} \cite{whw:02}, Woosley and Heger \cite{0004-637X-637-2-914}, and Heger \emph{et al.} \cite{Presupernova0004-637X-626-1-350}.  A detailed list of their properties (and a reference to the literature) is given in the upper box in \mbox{Table~\ref{radiustable}}. We also indicate in the table the references to the numerical results of the stellar progenitor models.

In addition to the models listed above, we examine other low- and solar-metallicity stars in the ZAMS mass range of (12 -- 35)\,M$_\odot$, also listed in the lower box in \mbox{Table~\ref{radiustable}}, to investigate some of our results' dependence on stellar mass and metallicity. Some of these models, which have ZAMS masses of $\lesssim30$\,M$_\odot$ will probably not create successful GRBs. Nevertheless, they can have choked relativistic jet activity that can be observed through neutrinos.

\section{High-Energy Neutrinos}
\label{sec:HENS}
\subsection{Production}
The variability in the output of the GRB's central engine results in
internal shocks within the jet, which accelerate electrons
and protons to high energies. Internal shocks can occur
even when the relativistic jet is still propagating inside the star
\cite{2001PhRvL..87q1102M}. For jets inside the star,
reverse shocks can also occur
\citep{chokedfromreverseshockPhysRevD.77.063007} at the head of the
jet, which can also accelerate electrons and protons to relativistic energies.

Relativistic electrons emit gamma rays through synchrotron or
inverse-Compton radiation. Relativistic protons interact with these
gamma rays ($p\gamma$), or with other non-relativistic protons ($pp$),
producing pions and kaons.
Photomeson interactions produce charged pions ($\pi^+$) through the
leptonic decay $p \gamma \rightarrow \pi^+$. Proton-proton
interactions produce charged pions ($\pi^{\pm}$) and kaons
($K^{\pm}$). Charged pions and kaons from these processes decay into
neutrinos through
\begin{equation}
\pi^{\pm}, K^{\pm} \rightarrow \mu^{\pm} + \nu_{\mu}(\overline{\nu}_{\mu}) 
\label{piondecay}
\end{equation}
Muons further decay to produce secondary neutrinos through, e.g., the
process $\mu^+ \rightarrow e^+ + \nu_e + \overline{\nu}_{\mu}$. However, if the synchrotron photon density is high enough, or in the presence of strong magnetic fields (i.e. for
smaller radii), they may immediately undergo radiative
cooling, giving a flux of lower energy neutrinos
\citep{HENAndoPhysRevLett.95.061103,HEN2005PhysRevLett.93.181101}.

The energy of charged mesons from both $p\gamma$ and $pp$ processes is
about $20\%$ of the proton's energy, while roughly $1/4^{th}$ of this
energy is given to $\nu_{\mu}(\overline{\nu}_{\mu})$
\citep{waxmanbachall}.  The energy of the produced neutrinos and
antineutrinos is therefore $\sim 5\%$ of the proton energy. The energies of the
photon ($\epsilon_{\gamma}$) and proton ($\epsilon_{p}$) in the
$p\gamma$ interaction need to satisfy the photo-meson threshold
condition of the~\mbox{$\Delta$-resonance}~\citep{waxmanbachall}
\begin{equation}
\epsilon_{\gamma}\epsilon_{p} \enspace \approx \enspace 0.2\Gamma^2 \, \mbox{GeV}^2
\label{photomeson}
\end{equation}
where $\Gamma$ is the Lorentz factor of the shock. Assuming Lorentz
factor $\Gamma \sim 300$ and observed $\gamma$ ray energy
$\epsilon_{\gamma} \sim 1$~MeV, one obtains a characteristic neutrino
energy of $\epsilon_{\nu} \sim 10^{14}$~eV \citep{waxmanbachall}.

For both $p\gamma$ and $pp$ processes, the energy spectrum of HENs is
determined mainly by the proton energy spectrum, and the optical
depths of the $p\gamma$ and $pp$ interactions. The distribution of the
proton energy $\epsilon_{p}$ in internal shocks, in the observer frame, is
$\mbox{d}^2N/(\mbox{d}\epsilon_{p}\mbox{d}t)\propto \epsilon_{p}^{-2}$, with a maximum
energy cutoff due to photo-pion losses. The cutoff energy depends on both jet
properties and the radius where the internal shocks occur.

The collisional or radiative nature of the internal shocks at high densities can be an impediment to Fermi acceleration. For mildly relativistic jets, particle acceleration happens less efficiently because the shock is somewhat spread out~\cite{Murase:2010cu}. However, the details of the acceleration process are still uncertain, and for many alternative acceleration scenarios this is not an issue~\cite{Arons:2002yj, Bucciantini:2007hy, Murase:2009pg}. On the other hand, neutrinos are the best probes for the physics responsible for the acceleration, and HEN observation should shed light on this aspect.

\subsection{Interactions and Oscillations}
\label{section:interaction}

HEN emission from internal shocks can commence at a distance
$r_s\approx\Gamma_{j}^{2}c\delta t\approx3\times10^9$~cm
\citep{2001PhRvL..87q1102M}, where $\Gamma_j=10$ is the jet Lorentz factor and $\delta t
= 10^{-3}\,$s is the jet variability time. Photo-pion losses determine the cut-off in the energy spectrum.

Whether and when HENs escape the star depends on the star's optical
depth to neutrinos, which depends on neutrino energy as well as where
neutrinos are emitted inside the star. HENs can only escape after the
relativistic jet reaches a low density region, beyond which the interaction of HENs with the stellar medium is negligible.

The effect of HEN interaction with matter on observable HEN emission
prior to the outbreak of the jet has been discussed previously, e.g., by Razzaque,
M\'esz\'aros, and Waxman \cite{PhysRevD.68.083001}.
Using a simplified model for the jet and progenitor star, Razzaque \emph{et al.} found that HEN
interaction is negligible if the jet gets close to the surface
(estimated as having roughly $\sim0.1$\,M$_\odot$ envelope mass over
4$\pi$ above the jet front). If the jet is deeper inside the star
(with overlying envelope material of $\sim1$\,M$_\odot$), they found
HEN interaction effects to be noticeable, especially for stellar
models that lost their hydrogen envelope. For this latter case, they
found that the neutrino optical depth becomes larger than unity for
neutrino energies $\epsilon_{\nu}\gtrsim2.5\times10^{5}\,$GeV. Horiuchi and Ando
\citep{chokedfromreverseshockPhysRevD.77.063007} also mention HEN
interactions for jets within the stellar envelope. They find that only
neutrinos with energies less than $\epsilon_\nu<10^2\,$GeV can escape a
progenitor star from~$r\approx10^{10}$~cm.


The interaction probability of HENs increases with their energy, and
can be non-negligible if a neutrino beam were to travel through large
quantities of dense matter, as is the case for neutrinos produced
inside massive stars. The mean free path of HENs is determined by inelastic scattering
processes, with neutrino-nucleon interactions ($\nu N$ or
$\overline{\nu} N$) being the major determinant of the optical depth
\citep{crosssectionPhysRevD.58.093009,flavoroscillation2009arXiv0912.4028R}. At
the relevant neutrino energies, the interaction cross sections are
approximately the same for all neutrino flavors
\citep{flavoroscillation2009arXiv0912.4028R}, therefore we treat
the cross section to be flavor independent (see \citep{flavoroscillation2009arXiv0912.4028R}). Electron-antineutrinos are the exception to the above, as the interaction of
electron-antineutrinos and electrons ($\overline{\nu}_ee$) becomes the
dominant effect around the resonant neutrino energy
$\epsilon_{\overline{\nu}_e} \approx 6.3\times10^6\,$GeV
\cite{crosssectionPhysRevD.58.093009}. We shall neglect this exception as
it only results in the attenuation of a small fraction of HENs,
i.e. the radius at which neutrinos of a given energy can first escape
does not change.


To obtain HEN interaction lengths, we adopt the neutrino cross
sections obtained by Gandhi et
al. \citep{crosssectionPhysRevD.58.093009} (other calculations give
similar results; see, e.g., \cite{crosssectionHill1997215}). We
additionally take into account nuclear effects (i.e. that both free
and bound nucleons are present) calculated by Pena~et~al.~\citep{2001PhLB..507..231C}. Gandhi \emph{et al.} \citep{crosssectionPhysRevD.58.093009} calculated
the cross sections for charged and neutral currents, the sum of these
two giving the total cross section $\sigma_\nu$. For the
range of neutrino energies of interest here, neutrino cross sections
are, to a good approximation, flavor invariant
\citep{crosssectionPhysRevD.58.093009,flavoroscillation2009arXiv0912.4028R}. We
note that the cross sections for neutrinos and antineutrinos are
somewhat different.

The neutrino-nucleon interaction cross section
$\sigma_\nu(\epsilon_\nu)$, obtained numerically by Gandhi~et~al. \citep{crosssectionPhysRevD.58.093009}, assumes that matter is in
the form of free nucleons. To take into account the presence of bound
nucleons (mostly helium), we approximate nuclear effects following
Castro~Pena~et~al.~\citep{2001PhLB..507..231C}. The ratio
of the total cross section $\sigma_\nu(A)$ corrected for nuclear
effects over the cross section $\sigma_\nu$ without correction
(free-nucleon case) decreases with energy and with atomic mass number
$A$. The effect is practically negligible (order of a few percent)
below a neutrino energy of $\epsilon_\nu \approx 10^{5}\,$GeV. Above
$\epsilon_\nu \approx 10^{5}\,$GeV, we approximate the energy
dependence presented in \citep{2001PhLB..507..231C} (see
Figure\,2 therein) with the
empirical function \be
\frac{\sigma_\nu(\epsilon_\nu,A)}{\sigma_\nu(\epsilon_\nu)} \approx
\left(\frac{\epsilon_\nu}{10^{5}\,\mbox{GeV}}\right)^{-\ln(A)/556}\,. \ee
Given the neutrino interaction cross sections, one can obtain the
interaction length (or mean free path) $\lambda_\nu$ via
\begin{equation}
\frac{1}{\lambda_\nu(\epsilon_\nu,r)} = \sum_A \rho(r)\omega_A(r) N_{\rm av}\sigma_\nu(\epsilon_\nu,A)  \enspace ,
\label{length}
\end{equation}
where $\rho(r)$ is the stellar density as a function of the radial
distance $r$ from the center of the star, $\omega_A(r)$ is the mass
fraction of elements with mass number $A$, $N_{\rm av}$ is the Avogadro
constant, and $\sigma_\nu(\epsilon_\nu)$ is the $\nu N$ interaction
cross section as a function of neutrino energy $\epsilon_\nu$ and
atomic mass number $A$. For antineutrino interaction length
$\lambda_{\overline{\nu}}$, the difference compared to $\lambda_{\nu}$
is the energy-dependent ratio of neutrino and antineutrino cross
sections.

Besides scattering, neutrinos also undergo flavor mixing due to neutrino flavor oscillations. After the neutrinos escape the star, the neutrinos travel a very long distance in space wherein the wave-packets of each mass eigenstate must separate. Therefore, neutrinos observed at Earth should be considered as incoherent superpositions of these mass-eigenstates. If the expected neutrino flavor ratio leaving the source is $\phi_{\nu_e}$:$\phi_{\nu_\mu}$:$\phi_{\nu_\tau} = 1$:$2$:$0$, neutrino oscillations from the source to the detector transform these ratios to 1:1:1 \citep{PhysRevD.68.093005}. If muon radiative cooling is significant, then only muon neutrinos are produced through the decay of kaons and pions, i.e. the flavor ratio at the source will be 0:1:0.

These ratios at the source and at the detector are 0:1:0 and 1:2:2, respectively \citep{PhysRevD.68.093005}. Other flavor ratios are possible and have been explored~\cite{Winter:2012xq}. As the energy and time profile of neutrino emission remains almost unchanged by oscillations, we do not consider these oscillation effects. We note that these ratios are modified by neutrinos oscillations inside the star that may be important for neutrino energies $\lesssim 10^4\,$GeV \cite{flavoroscillation2009arXiv0912.4028R}.

\subsection{Detection}

HENs traveling through Earth interact with the surrounding matter creating secondary particles, mostly muons. Cherenkov radiation from these muons is detected by neutrino detectors, which reconstruct direction and energy based on the detected photons \citep{likelihood_ratio_icecube}. While the IceCube detector \citep{2004APh....20..507A} has a threshold of $\sim10^2\,$GeV, Earth starts to become opaque to neutrinos with energies $\epsilon_\nu \gtrsim 10^6$~GeV, decreasing the neutrino flux that reaches the vicinity of the detector after crossing Earth \cite{2004APh....20..429G}. So-called \emph{upgoing} neutrinos (i.e. neutrinos whose trajectory crosses Earth before reaching the detector) with energies above $\sim10^6$~GeV are practically undetectable, as most of them are absorbed before reaching the detector (note that this is true for muon-neutrinos; tau-neutrinos can penetrate Earth even at higher energies \cite{2004APh....20..429G}). For upgoing muons this sets the observable energy window at $(10^2-10^6)\,$GeV.
\emph{Downgoing} and \emph{horizontal} neutrinos (i.e. neutrinos whose trajectory reaches the detector without crossing Earth) are detectable at practically any energy \cite{2004APh....20..429G}. The disadvantage of such directions is the much higher background noise from atmospheric muons that are, for these directions, not filtered out by Earth. Due to this high background, most HEN analyses only consider upgoing HEN events. The only exception is extremely-high energy neutrinos ($\gtrsim10^6\,$GeV), as atmospheric muons seldom reach such high energies.

There are several currently operating HEN telescopes, e.g.,
IceCube \citep{2004APh....20..507A}, a km$^3$ detector at the South
Pole, and \textsc{Antares} \citep{Antares2010arXiv1002.0754C} in the
Mediterranean sea. Both \textsc{Antares} and a third detector at
the lake Baikal \citep{Avrorin2011S13} are planned to be upgraded into
km$^3$ telescopes \citep{KM3NeTdeJong2010}. These ${\rm km}^3$-scale detectors observe HEN events and measure their energy, time, and arrival directions. While uncertainties remain in the emission mechanism and expected source flux \cite{2012Natur.484..351A}, depending on the model, these ${\rm km}^3$-scale detectors may observe $(1-100)$ HEN events in the energy range $(10^2-10^6)\,$GeV, for a typical GRB at $10\,{\rm Mpc}$.

The uncertainty of muon energy reconstruction in neutrino telescopes is $\sim0.3$ in $\log_{10}(\epsilon_\nu/{\rm GeV})$
\citep{Ahrens2004169,2003ICRC....3.1329R,PhysRevD.83.012001}, becoming somewhat larger \citep{PhysRevD.83.012001} for partially contained muons of greater energies. Timing resolution is expected to be on the nanosecond level \cite{IceCubeAchterberg2006155,2009ITNS...56.1141K}. Angular pointing is expected to be $\lesssim 1^\circ$ at these energies, which may allow identification of the source independently. Although we will not perform a detailed simulation of the observable signal, we shall keep in mind these experimental parameters.

\section{Calculations and Results}

In this section, we characterize the effect of neutrino interactions on the observable HEN energies from jets. This effect can modify the neutrino flux from choked jets as well as from precursor neutrino emission for successful jets. We consider internal shocks for the calculation of the temporal structure and derive the energy-dependent onset time and duration for HEN emission.

\subsection{Optical Depth for High-Energy Neutrinos}
\label{sec:optical}

\begin{figure}
\begin{center}
\resizebox{0.5\textwidth}{!}{\includegraphics{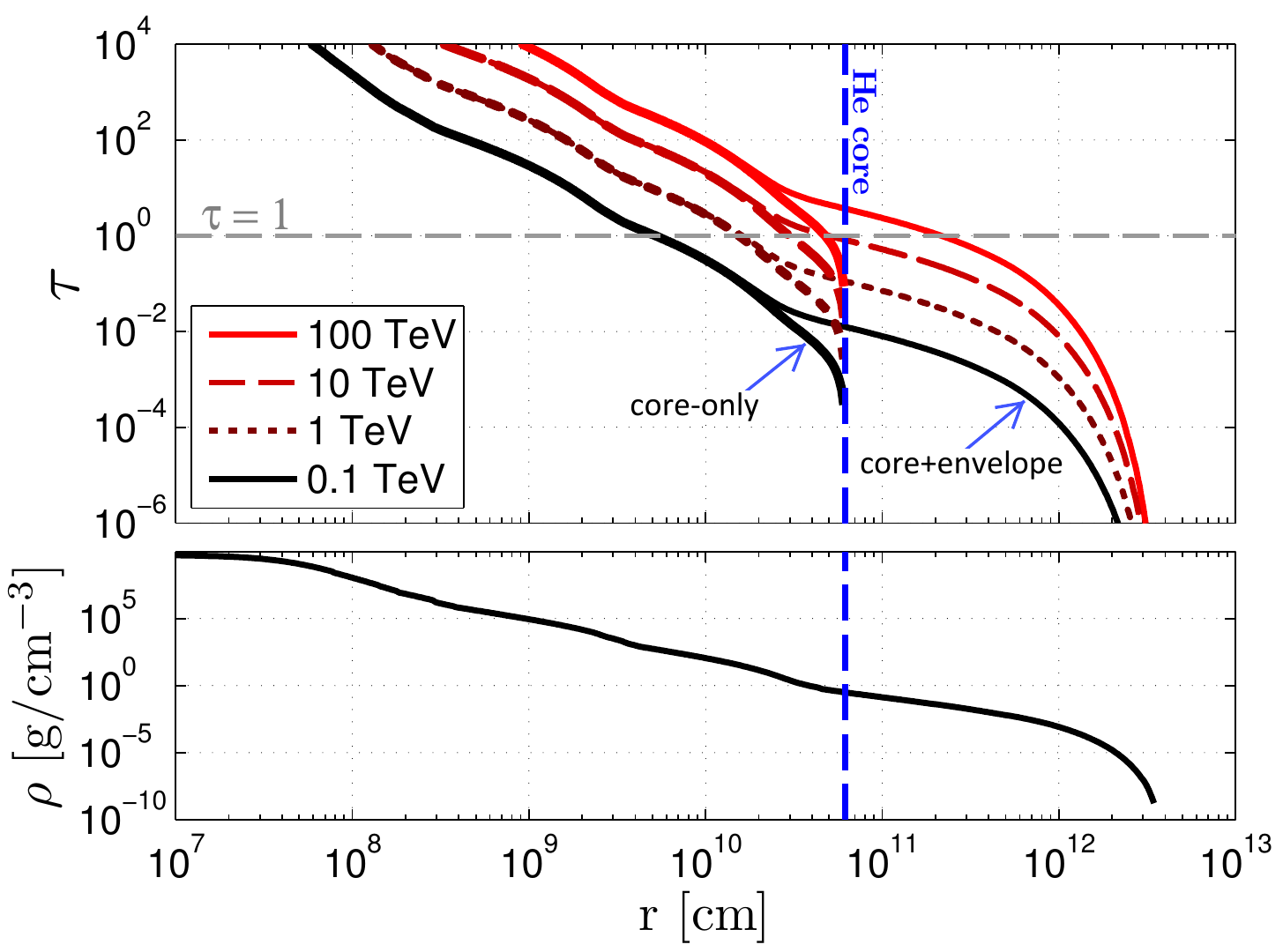}}
\end{center}
\caption{\textbf{(top)} High energy neutrino optical depths as a function of
  distance from the center of the star for different neutrino energies
  for a ZAMS 15\,M$_\odot$ star with low metallicity.  A vertical
  dashed line shows the radius of the helium core. As a comparison, we
  show the optical depth of the helium-core-only case indicated with
  thick lines. The horizontal dashed line shows $\tau =1$. Above this line, the stellar envelope is opaque to neutrinos. \textbf{(bottom)} The star's mass density as a   function of distance from the center.} \label{Henvelope}
\end{figure}

Given the mass distribution $\rho(r)$ in a star, one can employ the
expression for the neutrino mean free path described in the previous
section to calculate the HEN optical depth of the star for a given
distance from the center. We are interested in the innermost radius at
which neutrinos can escape from the star.

The optical depth $\tau$ of the star at a distance $r_0$ from its
center, towards neutrinos that are produced at
$r_0$ and are moving radially outward, is
\begin{equation}
\tau_\nu(\epsilon_\nu,r_0) = \int_{R}^{r_0}\frac{1}{\lambda_\nu(\epsilon_\nu,r)}\mbox{d}r
\label{equation:tau}
\end{equation}
where $R$ is the stellar radius. We approximate neutrino absorption such that neutrinos with energy $\epsilon_\nu$ cannot escape from below a \emph{critical radius} $r_\nu$ for which $\tau_\nu(\epsilon_\nu,r_\nu) \approx 1$.

We calculated the optical depths for the considered massive stellar
models (see Table \ref{radiustable}) as functions of radial distance
$r$ from the center of the star, as well as neutrino energy
$\epsilon_\nu$. In Figure\,\ref{Henvelope}, we exhibit the representative behavior of the neutrino optical depth as a function of
distance from the center for a stellar model with $M=15$\,M$_\odot$
ZAMS mass and low metallicity. We can see that the hydrogen envelope of the ZAMS
$M=15$\,M$_\odot$ star with low metallicity, for most relevant
energies, is transparent to neutrinos. We also see that, for this
stellar model, the helium core becomes opaque to neutrinos with all
depicted energies around $\approx 10^{10}$cm.

\begin{figure}
\begin{center}
\resizebox{0.5\textwidth}{!}{\includegraphics{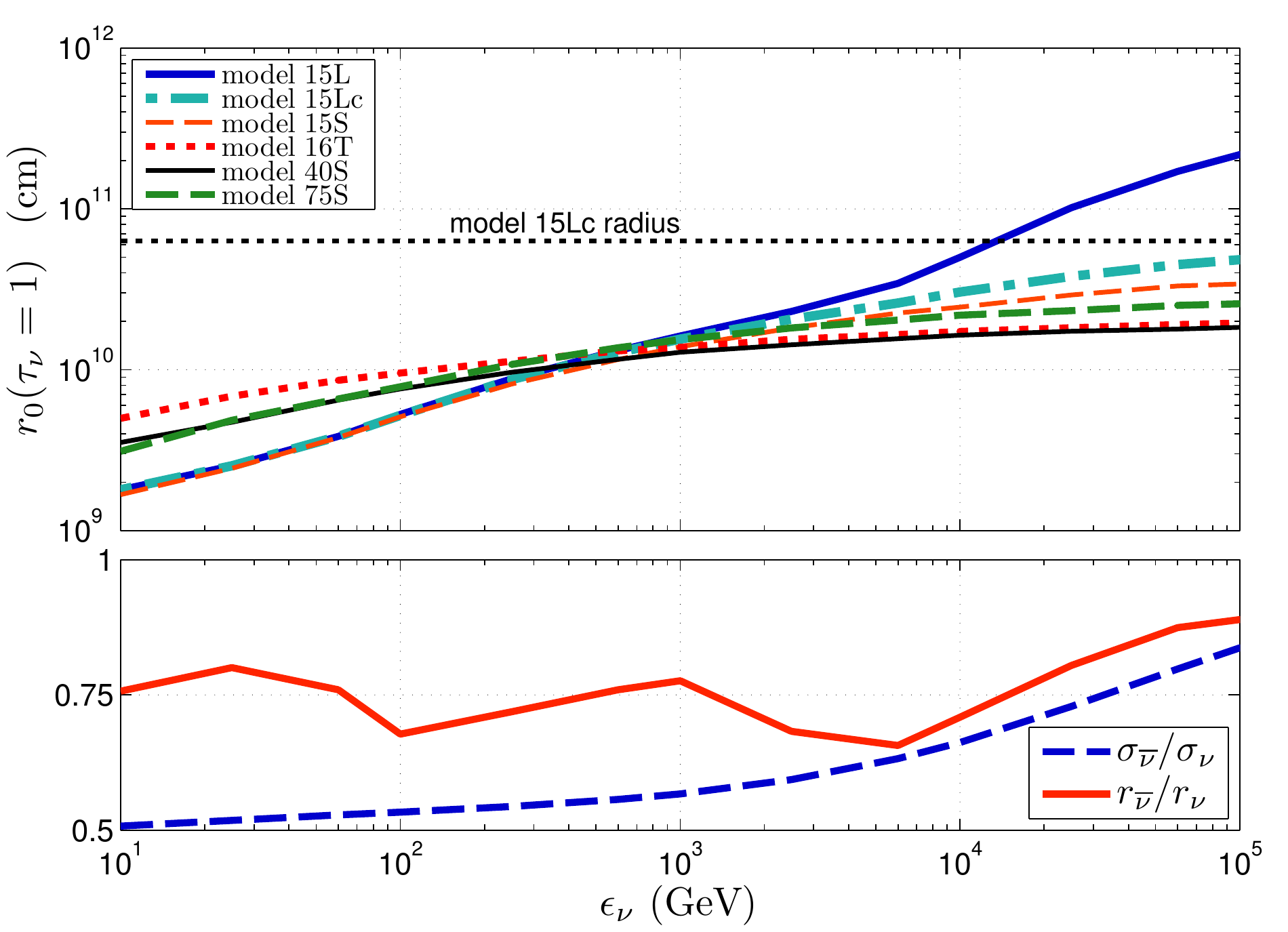}}
\end{center}
\caption{\textbf{(top)} Critical radius $r_\nu$ as a function of neutrino energy for the considered progenitor models (see Table \ref{radiustable}). The horizontal dashed line shows the pre-supernova helium-core radius for the ZAMS 15\,M$_\odot$ star with low metallicity (model 15Lc). \textbf{(bottom)} Ratio of cross sections for neutrinos and antineutrinos $\sigma_{\overline{\nu}} / \sigma_{\nu}$ (data taken from \citep{crosssectionPhysRevD.58.093009}), and the obtained ratio of the critical radii for neutrinos and antineutrinos, as the function of neutrino energy, for the ZAMS 20\,M$_\odot$ star with low metallicity.}
\label{opacity}
\end{figure}

To obtain $r_\nu$ as a function of neutrino energy, we inverted Equation \ref{equation:tau} to derive the critical radii.
In Figure\,\ref{opacity}, we present the critical radii for neutrinos. In the lower panel, we present the ratio of critical radius of antineutrinos and neutrinos (for the representative $M=20$\,M$_\odot$ low-metallicity case). One can see that the neutrino and antineutrinos have very similar critical radii. The maximum difference between the two radii is about $25\%$, indicating that our results on critical radii are valid for antineutrinos as well.

\begin{figure*}
\begin{center}
\resizebox{1\textwidth}{!}{\includegraphics{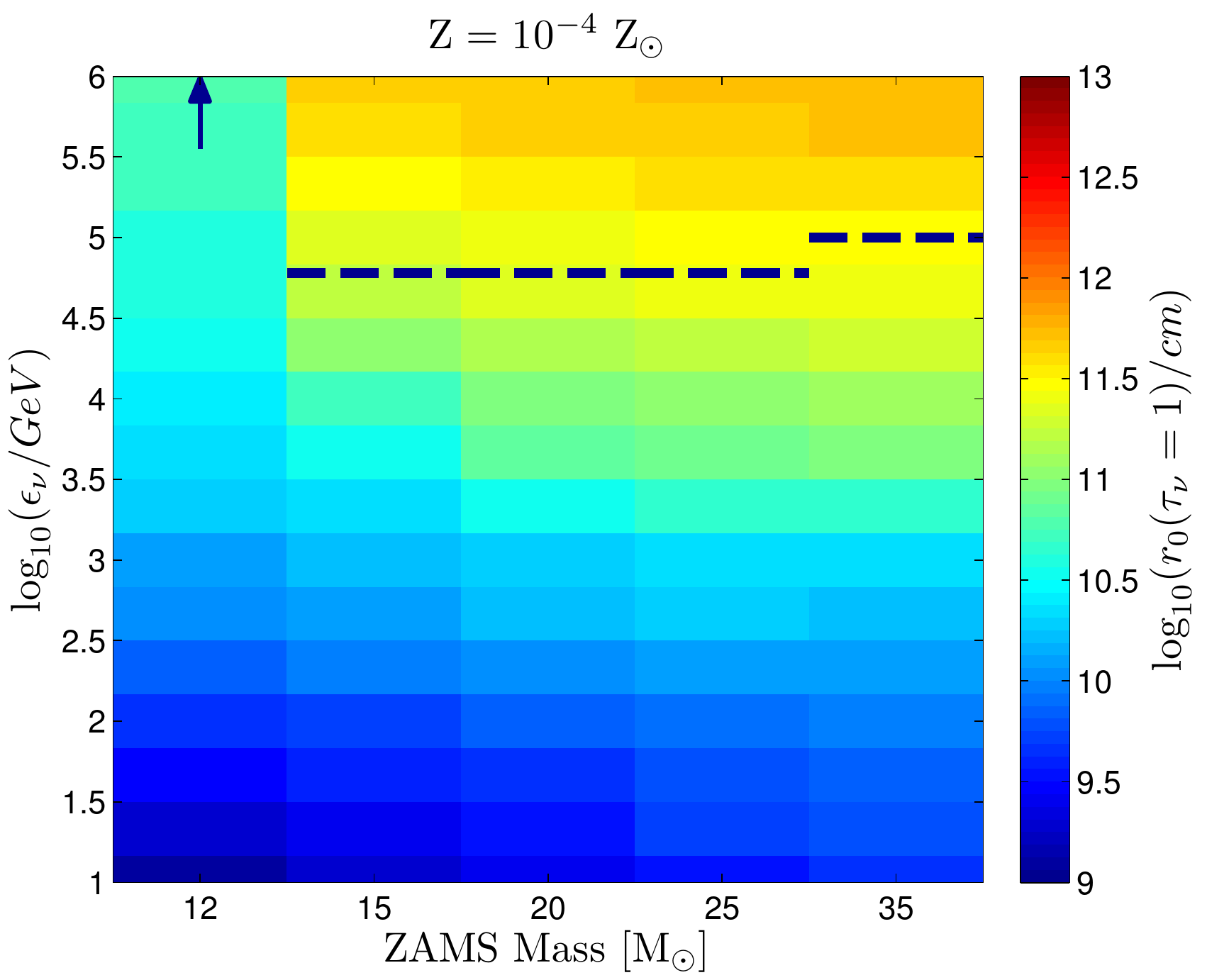}\quad\quad\includegraphics{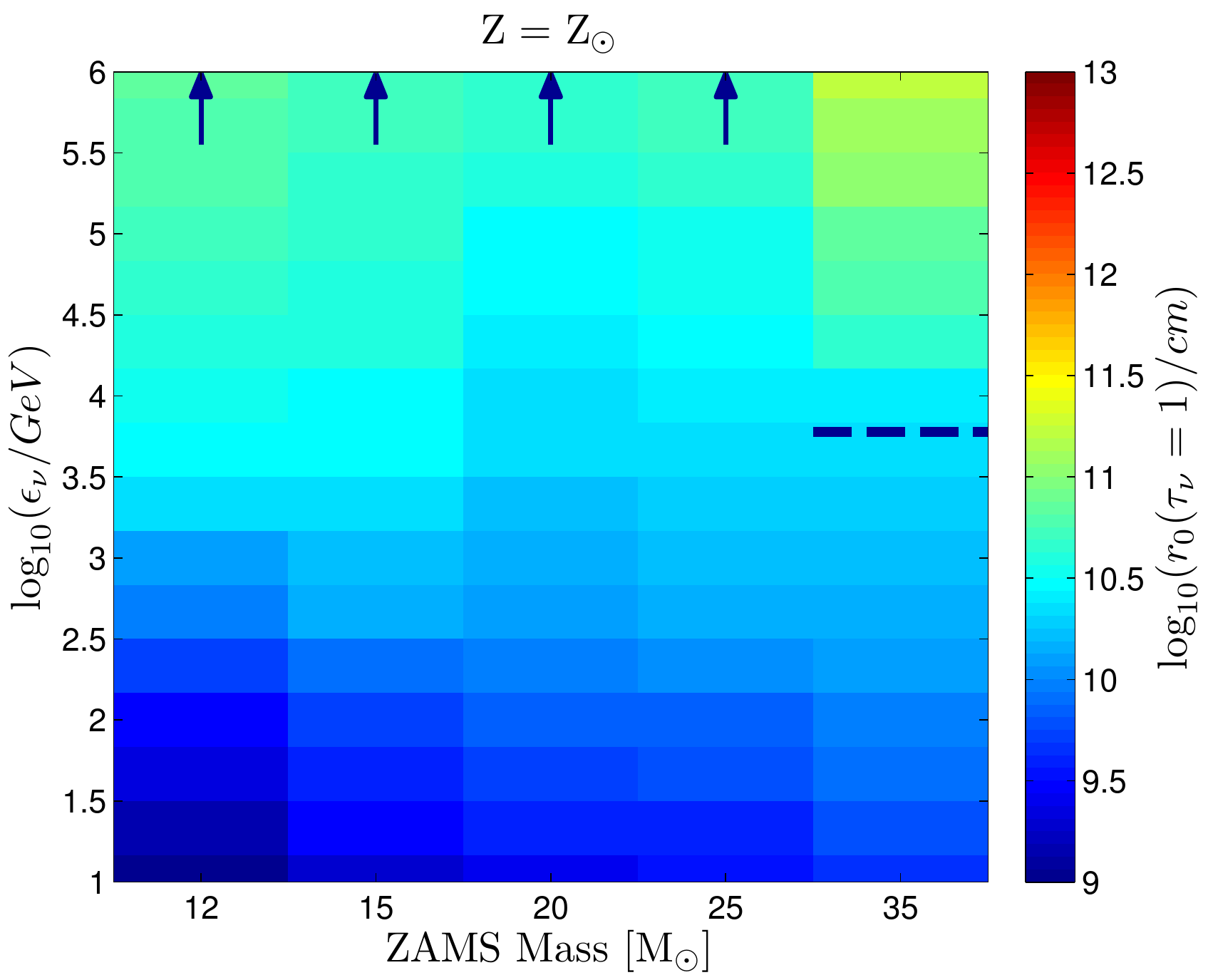}}
\end{center}
\caption{Critical radius $r_\nu = r_0(\tau_\nu=1)$ as the function of ZAMS stellar mass and neutrino energy $\epsilon_\nu$ for low-metallicity \textbf{(left)} and
  Solar-metallicity \textbf{(right)} simulations by Woosley \emph{et al.} \citep{whw:02}, and Heger \emph{et al.} \citep{Presupernova0004-637X-626-1-350}. The horizontal dashed lines show, for each ZAMS mass, the energy for which the critical radius is equal to the pre-supernova helium core radius (if all neutrinos with $\epsilon_{\nu} < 10^6\,$GeV can escape the core, the horizontal dashed line lies above the shown parameter space; this is indicated with arrows pointing upwards). Consequently, neutrinos with energies above the dashed lines cannot escape if produced inside the helium core. Note that, as shown for an example in Figure\,\ref{Henvelope}, the presence of the hydrogen envelope has only small effect on this threshold energy.} \label{figure:neutrinospheres}
\end{figure*}

We have investigated the dependence of the above results on the mass and metallicity of the progenitor. To do so, we repeated the above exercise for models with low and solar metallicity, from (12-35)$\,$M$_\odot$ listed in Table \ref{radiustable}, and plotted the critical radius as a function of energy and ZAMS mass, separately for low and solar metallicity. A contour plot indicating the effect of ZAMS stellar mass on $r_\nu$ for different metallicities and neutrino energies is shown in Figure\,\ref{figure:neutrinospheres}. The critical radius that corresponds to the supernova-progenitor's helium core is indicated with dashed horizontal lines (if all neutrinos with $\epsilon_{\nu} < 10^6\,$GeV can escape the core, the horizontal dashed line lies above the shown parameter space; this is indicated with arrows pointing upwards).

Figure\,\ref{figure:neutrinospheres} shows that the critical radius lies within the stellar helium core for neutrino energies of $\epsilon_\nu\lesssim10^5$ GeV for practically all massive progenitors. Consequently, the most relevant neutrino energy range for observations will become observable somewhere beneath the stellar core, making neutrino observations from choked-GRBs relevant. The results further indicate that $r_\nu(\epsilon_{\nu})$ is determined predominantly by the pre-supernova mass (and not the ZAMS mass) of the progenitors.

Beyond the optical depth of the stellar envelope, here we examine the possibility of neutrino absorption by the jet itself and the jet head. For the fiducial values of our model and the shock radius (see below; luminosity $L_{iso}=10^{52}\,$erg$\,$s$^{-1}$, jet gamma factor $\Gamma_j=10$, and shock radius $r_s=3\times10^{9}\,$cm), the density of the jet [see Eq. (\ref{equation:nj})] at $r_s$ in the observer frame is $\rho\approx0.3\,$g$\,$cm$^{-3}$. This density is negligible compared to the stellar density $\rho(r=3\times10^9\,\mbox{cm})\approx3\times10^3\,$g$\,$cm$^{-3}$ (model 15L), therefore the jet itself will not play a role in neutrino absorption.

To estimate the density of the jet head, we take stellar density $\rho(r=3\times10^9\,\mbox{cm})\approx3\times10^3\,$g$\,$cm$^{-3}$ (model 15L), and jet head Lorentz factor $\Gamma_h=0.1L_{52}^{1/4}r_{10}^{-1/2}\rho_{3}^{1/4}\rightarrow 0.13$ for the fiducial values \cite{2001PhRvL..87q1102M}. While the density of the jet head is significantly smaller than the stellar density, the shocked stellar density will be a few times greater than the stellar density \cite{chokedfromreverseshockPhysRevD.77.063007,2000ApJ...531L.119A}. Nevertheless, the thickness of this shocked region is small compared to the stellar radius. Taking the simulation of Aloy \emph{et al.} \cite{2000ApJ...531L.119A} as an example, the shocked stellar region at its peak is $\sim2.5\times$ greater than the stellar density when the jet head is at $6\times10^9\,$cm, and the thickness of the jet head is $8\times10^8\,$cm. With such jet head at $6\times10^9\,$cm, the stellar column depth increases by $\sim20\%$ (using model 15L). While such shocked stellar region does not change neutrino absorption substantially compared to the rest of the star, it further increases the role of neutrino absorption in the observable neutrino spectrum. Further, for faster advancing jets, the jet carries more matter in front of it as it takes time for matter to flow sideways \cite{2003MNRAS.345..575M}, making the shocked head even denser (see also \cite{chokedfromreverseshockPhysRevD.77.063007}). For simplicity, in the results below we neglect the absorption due to the shock region.

\subsection{Temporal Structure of Jets}
\label{sec:temporal}

Jets have been studied numerically and analytically, and a detailed understanding has been developed; see, e.g., references in \citep{chokedfromreverseshockPhysRevD.77.063007}. For our work, we use the semi-analytic method of Horiuchi and Ando \citep{chokedfromreverseshockPhysRevD.77.063007} to calculate the velocity of the jet head advancing inside the star in order to characterize the temporal structure of HEN emission. This simple treatment assumes a constant jet opening angle, and will suffice to illustrate our point, but in the future this may be improved through more detailed numerical modeling of the jet morphology for individual progenitors.

Horiuchi and Ando consider the
propagation of a relativistic jet with Lorentz factor
$\Gamma_j\gg1$. As the head of the jet advances through stellar matter
with Lorentz factor $\Gamma_h$, a reverse and forward shock occur. The
reverse shock decelerates the head, while a forward shock accelerates
the stellar material to $\Gamma_h$. In the following, we use the
subscripts j (jet, unshocked), h (jet head, shocked), s (stellar,
shocked), and ext (stellar, unshocked) to denote quantities at
different regions in and around the jet. Using this notation, the
evolutions of the two shocks at the jet head are governed by the
following equations \cite{Blandford:1976uq, Sari:1995nm}:
\begin{align}
e_s/n_sm_pc^2& =\Gamma_h-1,&
n_s/n_{ext}& =4\Gamma_h+3,\\
e_h/n_hm_pc^2& =\overline{\Gamma}_h-1,&
n_h/n_j& =4\overline{\Gamma}_h+3,
\label{equation:shock}
\end{align}
where $m_p$ is the proton mass, $c$ is the speed of light, and $n_i$
and $e_i$ are the particle density and internal energy measured in the
fluids' rest frames. Lorentz factors are measured in the lab frame,
except $\overline{\Gamma}_h= \Gamma_j\Gamma_h(1-\beta_j\beta_h)$ that
is measured in the jet's comoving frame ($\beta_ic$ is the
velocity). For a jet with constant opening angle, the jet particle
density at radius $r$ in the jet frame is
\begin{equation}
n_j(r) = \frac{L_{iso}}{4\pi r^2\Gamma_{j}^2m_pc^3}
\label{equation:nj}
\end{equation}
where $L_{iso}$ is the isotropic-equivalent jet luminosity. At the jet
head, the shocked jet head and shocked stellar matter are separated by
a contact discontinuity, and are in pressure balance. Equating their
pressures ($p_h=e_h/3$ and $p_s=e_s/3$, respectively) and using
Equations (\ref{equation:shock}) and~(\ref{equation:nj}), we arrive at
\begin{equation}
\frac{n_j}{n_{ext}} = \frac{(4\Gamma_h+3)(\Gamma_h-1)}{(4\overline{\Gamma}_h+3)(\overline{\Gamma}_h-1)}
\label{equation:jetpropagation}
\end{equation}
We numerically solve Equation (\ref{equation:jetpropagation}) using the
appropriate stellar particle number density $n_{ext}=n_{ext}(r)$ of the considered stellar models (see Table
\ref{radiustable}) to obtain the jet propagation velocity as a
function of radius. We took the terminal Lorentz factor of the jet head to be $\Gamma_j$ in the limit of zero density. Similarly to Ref. \cite{chokedfromreverseshockPhysRevD.77.063007}, we find that the velocity of the jet head is practically
independent of the gamma factor $\Gamma_j$ of the jet, and increases
with the isotropic equivalent energy $L_{iso}$ of the jet (in the
relativistic limit, the gamma factor of the jet head
$\Gamma\propto\Gamma_{j}^{1/4}$~\citep{2001PhRvL..87q1102M}).

\begin{figure*}
\begin{center}
\resizebox{1\textwidth}{!}{\includegraphics{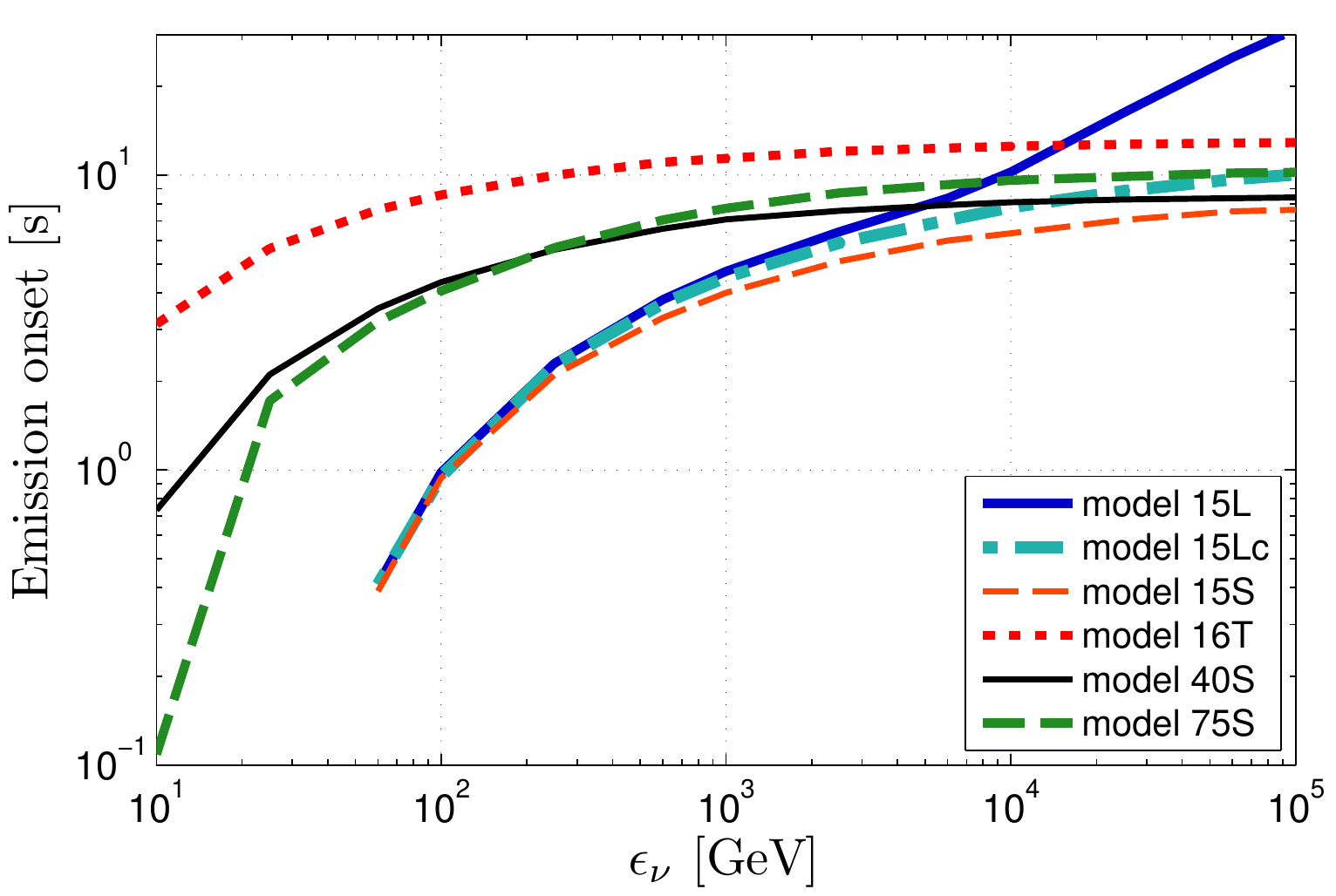}\includegraphics{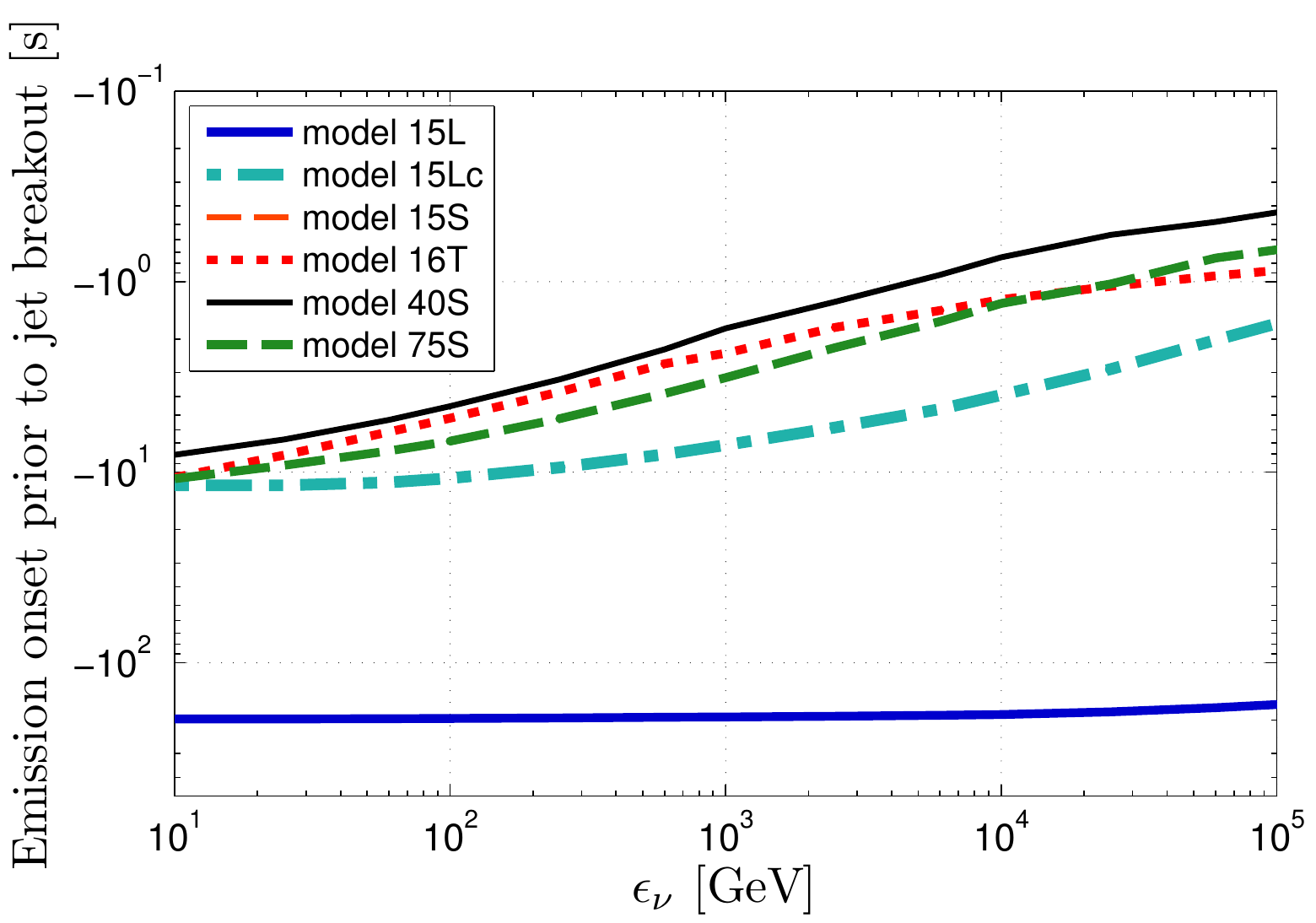}}
\end{center}
\caption{\textbf{(left)} Onset of observable HEN emission measured from the time when HEN production commences ($t_e(\epsilon_\nu) - t_0$), as a function of neutrino energy, for different stellar progenitors. \textbf{(right)} Time of jet breakout measured from the time of the onset of observable HEN emission ($t_{br} - t_e(\epsilon_\nu)$), as a function of neutrino energy, for different stellar progenitors. The calculations are carried out for the stellar models in Table \ref{radiustable} with jet energy $L_{iso}=10^{52}\,$erg$\,$s$^{-1}$ and jet Lorentz factor $\Gamma_{j}=10$.}
\label{figure:emissiononset}
\end{figure*}

\subsection{Energy-dependent Emission Onset Time}
\label{sec:onsets}

The skewed neutrino emission due to stellar neutrino-opacity can be characterized through the temporal structure of observable emission. Let $t_0$ be the time when the relativistic jet reaches the shock radius $r_s\approx\Gamma_{j}^{2}c\delta t\approx3\times10^9\,$cm. At this radius, internal shocks can form, causing HEN production to commence \cite{waxmanbachall}. Neutrinos that are produced while the jet is still beneath the stellar surface may only be able to escape from the star through the envelope at a later, energy-dependent ``escape" time $t_e(\epsilon_\nu)$, where $\epsilon_\nu$ is the energy of the neutrino. At $t_e(\epsilon_\nu)$ the jet has advanced far enough so the remaining envelope and the jet itself is no longer opaque to neutrinos with energy $\epsilon_\nu$. More specifically, we take $t_e(\epsilon_\nu)$ to be the time when the jet reached the critical radius $r_\nu$ for which $\tau_\nu(\epsilon_\nu,r_\nu) = 1$.

In Figure\,\ref{figure:emissiononset} (left), we show $t_e(\epsilon_\nu) - t_0$ (observer frame) as a function of $\epsilon_\nu$ for different stellar progenitors, which we calculated assuming a mildly relativistic ($\Gamma_{j}=10$) jet with $L_{iso}=10^{52}\,$erg$\,$s$^{-1}$ output. We can see that as neutrinos with lower energies can easily escape through the stellar envelope, one finds $t_e(\epsilon_\nu \ll 100\,{\rm GeV}) - t_0 \approx 0$.

It is also interesting to compare the escape time $t_e(\epsilon_\nu)$ with the jet breakout time $t_{br}$ from the stellar envelope. 
The point of jet breakout was chosen to coincide with the jet reaching the radius at which the simulated stellar progenitor models end. This corresponds to a sharp drop in matter density, dropping below $10^{-10}\,$g$\,$cm$^{-3}$ for models 15L and 16T, and dropping below $10^{-4}\,$g$\,$cm$^{-3}$ for models 40S and 75S, while the simulation ends at $10^{13}\,$cm for model 15S. Due to the low densities at the boundary of the simulated progenitors, the results should be robust to the specific choice of jet breakout radius.
We note that the bulk of the gamma-ray emission from the jet may be shortly delayed compared to the breakout due to dissipation within the jet until the jet advances to a distance of $\gtrsim10^{13}\,$cm \cite{2005ApJ...628..847R,2006RPPh...69.2259M,PrecursorLi,2011ApJ...733L..40M}. The GRB will become detectable within a few seconds after the jet head leaves the helium core \cite{2001ApJ...556L..37M}.

Figure\,\ref{figure:emissiononset} (right) shows the time it takes for the jet to break out of the star from the point from which neutrinos can first escape (i.e. $t_{br} - t_e(\epsilon_\nu)$), as a function of $\epsilon_\nu$ for different stellar progenitors. As before, a mildly relativistic ($\Gamma_{j}=10$) jet with $L_{iso}=10^{52}\,$erg$\,$s$^{-1}$ output has been used. A disadvantage of using $t_{br}$ for comparison is that it is only available for successful jets, and therefore it cannot be used to characterize, e.g., choked~GRBs.

As the figure shows, when the jet is close to the center of the
progenitor, only low energies neutrinos can escape the star. Gradually, as the jet proceeds
outward, higher energy neutrinos become observable. Consequently, neutrinos escaping (and observed) earlier are expected to have lower energies, leading to a time dependent neutrino energy distribution, with the average energy increasing with time. Such distribution may be indicated upon the detection of multiple neutrinos from a CCSN. The precise relationship of the energy-dependent onset times and emission durations of HENs can encode information about the progenitors' density profile and composition.

\subsection{Dependence on source parameters}

The results presented above were calculated for sources with mildly relativistic jets ($\Gamma_{j}=10$) with $L_{iso}$ luminosity, and with jet variability of $\delta t$. From these parameters, $\Gamma_{j}$ and $\delta t$ determine the shock radius $r_s$  (see Sec.\,\ref{sec:onsets}). For greater $\Gamma_{j}$ and $\delta t$, neutrino production will commence only at a greater radius, changing the lowest radius and earliest time from which neutrinos are observable. The jet luminosity $L_{iso}$ affects the jet head velocity, with greater (smaller) $L_{iso}$ corresponding to greater (smaller) velocity, which in turn change the time scale on Figure\,\ref{figure:emissiononset}; the jet head velocity is practically independent of $\Gamma_{j}$ and $\delta t$ (see also \cite{chokedfromreverseshockPhysRevD.77.063007}). These dependencies, however, do not affect our conclusions qualitatively.

\section{Interpretation of Energy and Time Structure of Neutrino Emission}
\label{sec:results}

\subsection{Strong-signal limit}

For a very nearby CCSN or GRB, one expects a modestly large HEN flux of perhaps $\gtrsim 100$ events. Although this possibility is rare, it does represent a rather lucrative opportunity. The highest energy neutrino detected from such a source at any time constrains the total matter content of the envelope above the jet head at that time. Therefore, one can take the opportunity to map the inner stellar density profile and the jet's velocity.

Comparison of the emission onset profiles of different stellar progenitor models in Figure\,\ref{figure:emissiononset} (left panel) shows that, with a high-enough HEN flux, the stellar models have distinguishable energy-time profiles (i.e. cutoff energies as a function of time). For example, for a 5s long precursor emission, the groups of models \{16T\}, \{40S,75S\} and \{15L, 15Lc,15S\} are distinguishable, while members of a given group have practically the same energy cutoff.

Another consequence of the radius-dependent neutrino energy cut-off is that neutrinos emitted by choked relativistic jets (e.g., choked
GRBs) will have an energy cutoff, resulting in lower average energy than jets that successfully break through to the surface of the
star (e.g., successful GRBs). Such a difference can possibly be indicated upon the detection of a sufficient number of neutrinos from a set of
CCSN with and without electromagnetic counterparts. The indication of a difference in average energies from successful and choked jets can
provide information, e.g., on the distance the choked jets advance before they stall, a possible indicator of the energy and/or baryon
content of the jets.

\subsection{Weak-signal limit}

Most observed astrophysical sources will only lead to ${\cal{O}}(1)$ detected neutrinos. Therefore, we find it more practical to consider the possibility that only a few neutrinos are detected.  In the following, we explore some ideas for recovering structural information of the progenitors using only a few observed HENs. Such recovery may not be conclusive for every detected source, but for specific detections where the energy and timing of detected neutrinos are favorable, it can provide important information about the source.

\subsubsection{HEN timing prior to time of jet breakout}

The time difference between the onset of the (energy-dependent)
observable HEN emission and the jet breakout can be used to constrain the possible
progenitors, potentially even with one detected HEN. If the observation
time of a detected neutrino and the time of the jet breakout differs more than what is
allowed for a progenitor model, the respective progenitor model can be ruled out or weakened.
For example, if one detects a HEN with reconstructed energy $\epsilon_{\nu}\approx10^3$~GeV
approximately $\sim$5~sec before the jet breakout, Figure\,\ref{figure:emissiononset} (right) suggests that one can
practically rule out models \{15S,16T,40S,75S\}, while models \{15L,15Lc\} are
possible progenitors. We note that it may be difficult to directly observe the time of jet breakout as the gamma-ray photosphere typically lies above the stellar surface.

\subsubsection{HEN relative timing}

It is also possible to constrain progenitors even if the jet does not
break out of the star. In this case, the time difference between at least two
observed neutrinos can be used to constrain possible progenitors by determining whether
the observed time difference is possible or likely for a given progenitor model and the reconstructed
neutrino energies.
For example, the observation of two neutrinos, both
with energies $\epsilon_{\nu}\approx10^3$~GeV with $\sim$10~sec time difference would rule
out all models but 15L out of those considered in Figure\,\ref{figure:emissiononset}, indicating the presence of a hydrogen envelope.

\subsubsection{Jet duration vs. onset of neutrino emission}

Neutrinos above $\epsilon_{\nu}\gtrsim10^3$~GeV for all models but 15L are emitted only in a small fraction of the time the jet spends within the progenitor. Since it is unlikely that jets are fine-tuned such that they stop just before breaking out of the star, the detection of HENs with $\epsilon_{\nu}\gtrsim10^3$~GeV from confirmed astrophysical sources with no EM counterparts would make it highly probable that the progenitors have kept their hydrogen envelope prior to explosion.

\subsubsection{Neutrinos and gravitational waves}

The coincident observation of gravitational-waves and HENs from a common source has far-reaching astrophysical implications \cite{PhysRevLett.107.251101,Baret20111}. The relation between the observed times of arrival of gravitational waves and HENs from a stellar core collapse could provide information on the stellar structure below the shock region of $\gtrsim 10^{10}\,$cm (complementary to information from HENs and EM radiation, which map the structure at and above the shock region).

The time relativistic jets take to cross the stellar envelope is likely comparable to or less than the observed duration of prompt gamma-ray emission (as the duration of the outflow fed by the central engine is unlikely to be fine-tuned to the envelope crossing time). Given the observed long-GRB durations of $\approx (10-100)\,$s and expected stellar progenitor densities at small radii, the jet is only likely to be able to cross the envelope within $\approx (10-100)\,$s if the stellar density significantly decreases along the rotational axis~\cite{2001ApJ...556L..37M}. Gravitational-wave emission from the stellar progenitor is likely connected to the onset of jet propagation (core-collapse and the formation of an accretion disk are both potential sources of gravitational waves \cite{2012PhRvD..85j3004B}). Consequently, comparing the time of arrival of gravitational wave signals and HENs from a GRB or supernova progenitor can provide information on stellar densities below the shock region, and may provide information on the development of jets (see \cite{2001ApJ...556L..37M}).

\subsubsection{Extremely high energy neutrino triggers}

Extremely high energy neutrinos of energies $\epsilon_\nu\gtrsim100\,$TeV may be emitted once the relativistic jet becomes sparse enough such that neutrino production is not suppressed by strong magnetic fields or high synchrotron-photon density \cite{PhysRevD.68.083001,chokedfromreverseshockPhysRevD.77.063007}. The emission of such extremely high energy neutrinos may commence only once the jet has advanced substantially, likely outside the helium envelope \cite{chokedfromreverseshockPhysRevD.77.063007}. These neutrinos will therefore be produced only \emph{after} most lower energy HENs are created while the jet is at lower radii (we note that the central engine of GRBs may show activity on a longer time scale than the prompt gamma-ray emission \cite{2005ApJ...631..429I}, therefore weaker, longer duration neutrino emission is also plausible). Processes other than internal shocks, such as the jet's interaction with the interstellar medium, can further produce so-called ultra high energy neutrinos with energies up to $10^{10}\,$GeV that may be of interest here \cite{2000ApJ...541..707W,Waxman:1999ai,UHEN2}. Hence, it may be interesting to search for $\gtrsim100\,$TeV neutrinos in coincidence with $\lesssim100\,$TeV neutrinos that arrived \emph{prior} to an extremely high energy neutrino up to tens of seconds.

While HEN detectors mainly use Earth as a shield from atmospheric muons, extremely high energy neutrinos with $\epsilon_\nu\gtrsim1\,$PeV are also absorbed before they can cross Earth, and thus can only be detected from downgoing and horizontal directions \cite{2004APh....20..429G}. Even though there is an abundant atmospheric muon background from these directions in the detector, these muons have lower energies; hence, astrophysical extremely high energy neutrinos are relatively easy to identify.

Based on the above model, a search could be performed for extremely high energy neutrinos in coincidence with $\lesssim100\,$TeV neutrinos that arrived from the same direction, prior to the extremely high energy neutrino up to tens of seconds. This search could be particularly interesting, as $\lesssim100\,$TeV neutrino data have not been utilized in searches for astrophysical neutrinos (nevertheless, these downgoing events are recorded and stored for~\mbox{IceCube}~\cite{Toscano2012}).


\section{Conclusions}
\label{sec:summary}

We investigated the opacity of massive pre-supernova stars to high energy neutrinos (HENs) in order to address the question: What do the times of arrival of detected high energy neutrinos tell us about the properties of their source? We investigated the effect of opacity on the observable HEN emission from both successful and choked jets. In particular, we have examined various zero-mass main sequence (ZAMS) stellar masses from (12-75)\,M$_\odot$ for low-metallicity non-rotating and stellar-metallicity rotating cases.

For the considered progenitors, we presented the energy-dependent critical radius from which HENs cannot escape. We found that the presence of the stellar hydrogen envelope has a negligible effect on the optical depth for neutrino energies of $\epsilon_\nu\lesssim10^5$ GeV, i.e. the most relevant energy range for HEN detection. The critical radius, however, largely varies with $\epsilon_\nu$ within the helium core, which has relevant consequences on observations.

The neutrino emission spectrum changes as the relativistic jet advances in the star. Considering mildly relativistic jets ($\Gamma_j\approx10$) and HEN production in internal shocks, the energy dependence of the onset of neutrino emission is shown in Figure\,\ref{figure:emissiononset}. Such time dependence can provide important information on the stellar structure. For instance, with observation of multiple HENs, the detected neutrino energies and times provide constraints on stellar density at different depths in the star. The energy of a detected precursor HEN can also provide information on the maximum time frame in which the relativistic jet will break out of the star and become observable.

We examined how neutrino interaction with dense stellar matter can be used to probe stellar progenitors. We investigated both the strong and weak-signal limits, i.e. when many or only a few neutrinos are detected, respectively. We demonstrated that under favorable conditions one can use the time difference between a precursor neutrino and the jet breakout to exclude some progenitor models. The relative times of arrival for multiple neutrinos may also be sufficient to exclude progenitor models, even with no observed electromagnetic emission. Additionally, the detection of HENs with energies above $\gtrsim 10^3$~GeV from choked GRBs makes it likely that the progenitor possessed a hydrogen envelope prior to explosion. Also, while the detection of HENs and electromagnetic signals can provide information on the stellar region at and above the shock radius, the coincident detection of HENs with gravitational waves \cite{PhysRevLett.107.251101} may be informative w.r.t. jet development and propagation below the shock radius. We proposed the use of extremely-high energy neutrinos ($\epsilon_\nu\gtrsim100\,$TeV) detected by cubic-kilometer neutrino detectors such as IceCube for searches for coincident \emph{downgoing} neutrinos. These extremely-high energy events probably arrive after other lower energy events that can be used in a coincident analysis. Downgoing neutrinos with lower energies are typically not used in searches due to the large atmospheric muon background from these directions and energies.

A future extension of this work will be the calculation of neutrino fluxes from different radii, similar to the calculations of Razzaque \emph{et al.} \cite{PhysRevD.68.083001}, who estimated the flux for two different radii, but without timing information. Such addition to the temporal structure of neutrino energies can provide a more detailed picture of not only the information in neutrinos about the progenitor, but also the likelihood of detecting neutrinos with given information content.

We note here that the uncertainty in the reconstructed energy and timing of HENs introduces uncertainty in the measurement of the onset of HEN emissions at the energies of the detected neutrinos. These uncertainties need to be taken into account when comparing emission models to observations. Additionally, there could be other factors, e.g., physics related to jet propagation, that we have treated schematically in this work, but could have similar impact on the temporal structure of HEN events. However, we have pointed out some generic features which should motivate future work that investigates experimental detectability of these features.

The results presented above aim at describing the interpretation of a set of detected high energy neutrinos from a collapsar event. One of the advantages of such interpretation is that it can be done independently of the emitted high energy neutrino spectrum inside the source, which greatly simplifies the understanding of observations.  An important direction of extending this work will be the calculation of the emitted neutrino spectrum inside the source. Obtaining this time-dependent spectrum would let us understand the probabilities of obtaining a set of neutrinos with specific energies and arrival times, which can be used to further refine the differentiation between progenitor scenarios. The calculation of emission spectra can be done similarly to the work of Razzaque \emph{et al.} \cite{PhysRevD.68.083001}, who calculated the observable spectrum for two specific radii for one stellar model. For accurate flux estimates, one must also take into account neutrino flavor oscillations in the stellar envelope, which are non-negligible for the relevant energy range \cite{flavoroscillation2009arXiv0912.4028R}. For precisely assessing the capability of differentiating between various stellar models, one will further need to compare the estimated source flux with the atmospheric neutrino background of neutrino detectors (see, e.g., \cite{2004PhRvD..70b3006B,1996PhRvD..53.1314A,PhysRevD.83.012001}).

\acknowledgments

We thank Christian Ott for collaboration in early stages. We specially thank John Beacom, Alessandra Corsi, Doug Cowen, Chad Finley, Shunsaku Horiuchi, Peter M\'esz\'aros and Kohta Murase for valuable discussions and feedback on the manuscript. I.B. and S.M. are grateful for the generous support of Columbia University in the City of New York and the National Science Foundation under cooperative agreements PHY-0847182.

\bibliographystyle{h-physrev}


\end{document}